\documentclass[11pt]{article}
\usepackage{amsmath,amsthm,latexsym,amssymb,amsfonts,epsfig,psfrag}
\usepackage{graphicx, color, calc}

\makeatletter \@addtoreset{equation}{section} \makeatother

\title{On background-independent renormalization of spin foam models}

\author{Benjamin Bahr$^1$\\
\small $^1$ II Institute for Theoretical Physics\\
\small University of Hamburg\\
\small  Luruper Chaussee 149\\
\small 22761 Hamburg, Germany
 }
\date{}

\begin{document}

\maketitle

\begin{abstract}
\noindent In this article we discuss an implementation of renormalization group ideas to spin foam models, where there is no a priori length scale with which to define the flow. In the context of the continuum limit of these models, we show how the notion of cylindrical consistency of path integral measures gives a natural analogue of Wilson's RG flow equations for background-independent systems. We discuss the conditions for the continuum measures to be diffeomorphism-invariant, and consider both exact and approximate examples.
\end{abstract}


\section{Introduction}

The application of renormalization techniques to quantum gravity theories has always been problematic. This is due to the inherent tension between the notion of scale-dependent physics of the renormalization group on the one side, and the role of background-independence of general relativity on the other side.

The renormalization group is a tool for dealing with a phenomenon that is inherent to many physical systems with a very large number of degrees of freedom, which are nonlinear, i.e.~exhibit nontrivial self-interaction. To describe such systems, it is convenient to not consider all degrees of freedom on an equal footing, but rather to order them according to some hierarchy. This way one considers only a finite and manageable amount of the system at a time. This hierarchy often runs along a length scale, introducing a notion of ``smaller'' and ``larger'' degrees of freedom. Often the dynamics on different scales are related to each other, in that collective effects of smaller degrees of freedom combine to larger ones. The non-linearity of the interaction between the small degrees of freedom influences the dynamics of the system on larger scales. This leads to a scale-dependence of the effective actions for the system, and this ``running of coupling constants'' is a key concept in Wilson's approach to the renormalization group \cite{Wilson:1971bg, Wilson:1973jj}.

On the other hand, one of the key concepts of general relativity is its general covariance, which makes the theory background-independent. In particular, diffeomorphisms of the space-time manifold act as gauge symmetries. Arguably, the corresponding quantum theory should possess the same symmetries. This makes an easy distinction of ``small'' and ``large'' degrees of freedom difficult, because there is no external geometry to measure any length with. Rather, because geometry itself is dynamical, spatial dimensions become observables, rather than parameters. The length scales themselves become degrees of freedom, making the applicability of usual renormalization techniques to quantum gravity theories difficult.\\[5pt]

In this article, we will consider the -- tentatively background-independent -- spin foam model approach to quantum gravity, which has been very successful in describing a path integral for quantum gravity in terms of quantized geometry \cite{Barrett:1997gw, Engle:2007wy, Freidel:2007py, Engle:2009ba, Baratin:2011hp}, \cite{Livine:2007vk, Freidel:2010aq}, see also \cite{Perez:2012wv} for a review. A crucial ingredient to spin foam models is an abstract discretization of space-time. We will show how working with \emph{embedded}, rather than abstract, discretizations provides not only a natural notion of continuum limit, it also results in a conceptually clear analogue of ``scale'' in these models, as well as a mathematically well-defined notion of renormalization group flow for the background-independent context.

\paragraph*{The plan of the paper:}

In section \ref{Sec:HolonomySFM} we will discuss an approach to spin foam models based on the KKL-formalism \cite{Kaminski:2009fm, Bahr:2010bs, Bahr:2012qj}, which works with embedded, rather than abstract, discretizations of space-time. As a result, there is a natural notion of continuum limit in the sense of a projective family. In order to construct the continuum path integral measure, the partial measures need to satisfy a strong condition, called \emph{cylindrical consistency}. We show how these conditions are precisely the analogue of Wilson's RG flow equation in the background-independent context, and discuss the physical interpretation in terms of configurations and measurements.

In section \ref{Sec:DiffInvariance} we will demonstrate that the condition for the
continuum path integral measures to be invariant under space-time diffeomorphisms, can be posed naturally within this context, and show how this is a very strong condition on the partial measures defined at each scale.

In section \ref{Sec:Example} we will discuss an easy example, where the RG flow can be solved completely and exactly, and demonstrate how the condition of diffeomorphism-invariance severely restricts the set of solutions, most of which spontaneously break background-independence.

In section \ref{Sec:Approximations} we will discuss several suggestions for how to derive approximate solutions in more complicated cases, and give an example with a quartic interaction term, which can be treated with these methods.

While appendix \ref{App:Properties} provides mathematical details for some of the constructions, in appendix \ref{App:Canonical} the connection to refined algebraic quantization techniques is discussed, to make contact to the canonical framework in the case that space-time has a boundary.

\section{Holonomy spin foam models}\label{Sec:HolonomySFM}


In this section we review the definition of (holonomy) spin foam models \cite{Bahr:2012qj}. These are generalizations of lattice gauge theories defined on arbitrary, irregular lattices. They encompass candidate models for quantum gravity such as the  Barrett-Crane \cite{Barrett:1997gw} and the EPRL-FK model \cite{Engle:2007wy, Freidel:2007py}, as well as versions of group- and tensor field theories \cite{Oriti:2006se, Gurau:2011aq}. Spin-net models, which are lower-dimensional analogues of spin foam models \cite{Dittrich:2011zh}, are very close on a technical level, so that many results are transferrable \cite{Dittrich:2012jq, Dittrich:2013uqe, Dittrich:2013voa}.

\paragraph*{2-complexes:}

Throughout the article, we will work in the semi-analytical category, which allows us to circumvent certain technical inconveniences which arise when working with smooth structures, while being more ``local'' than using analytic functions. So all maps, manifolds and connections are supposed to be semi-analytic \cite{Semi01, Semi02, Braukhoff:2012} in what follows.\\[5pt]

To begin with we fix a manifold $\mathcal{M}$, which is supposed to be the model for space-time on which the fields are defined. For the time being $\mathcal{M}$ is supposed to be closed. The case of compact $\mathcal{M}$ with nonempty boundary will be treated in appendix \ref{App:Canonical}. The quantum theory that is developed then describes the fluctuation of connections of some type on $\mathcal{M}$.

The key concept of the approach will be that of a $2$-complex. By a $2$-complex we mean an abstract complex $\Gamma$ embedded in $\mathcal{M}$, where all 1- and 2-cells (``edges'' and ``faces'') are oriented. Wlog all 0-cells (``vertices'') are negatively oriented. For an edge $e\in \Gamma^{(1)}$ and a face $f\in\Gamma^{(2)}$ which contains $e$ as part of its boundary, define by $[e,f]=\pm1 $, depending on whether the orientation induced on $e$ by the given one on $f$ coincides or is opposite to the given one on $e$. Similarly $[v,e]=\pm 1$ whenever $e$ is incoming / outgoing w.r.t.~the vertex $v$.

\begin{figure}[hbt!]
\centering
\def\svgscale{0.75}
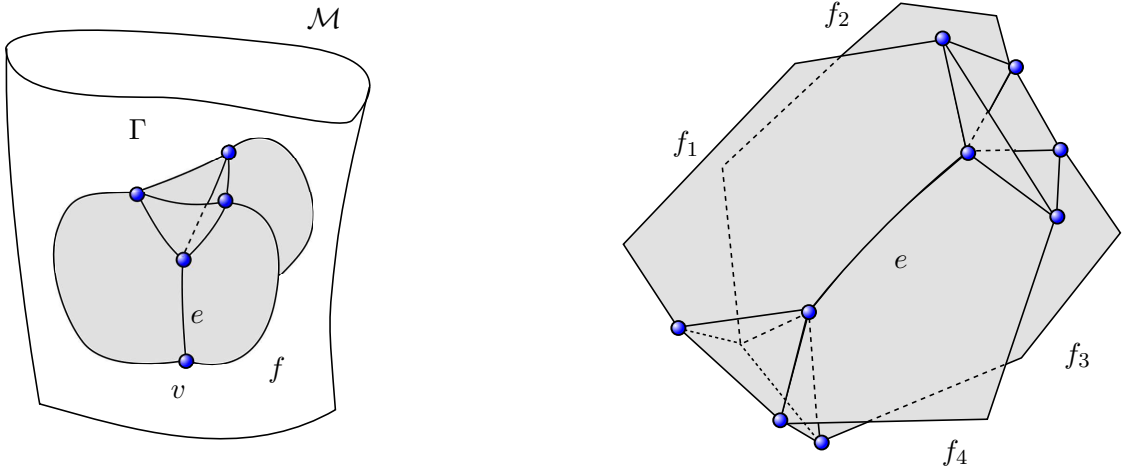
\caption{Left: Embedded $2$-complexes $\Gamma$ consist of vertices $v$, edges $e$ and faces $f$. Right: A configuration is as many group elements $h_{ef}$ per edge $e$ as there are faces $f$ touching it. In this case there are four group elements associated to the edge $e$: $h_{ef_1}$, $h_{ef_2}$, $h_{ef_3}$, and $h_{ef_4}$. }\label{Fig:Figure01}
\end{figure}

%
%

We note that the set of all such embedded $2$-complexes possesses an ordering relation: we write
\begin{eqnarray}\label{Eq:OrderingRelation}
\Gamma\,\leq\,\Gamma'
\end{eqnarray}

\noindent if and only if every cell of every dimension in $\Gamma$ is composed (disregarding orientation) of cells of the appropriate dimension in $\Gamma'$. Note that this depends crucially on the embedding of $\Gamma$, $\Gamma'$ in $\mathcal{M}$. In particular, every vertex in $\Gamma$ is necessarily a vertex in $\Gamma'$.

\paragraph*{Configuration spaces:}

Fix a compact Lie group $G$. For each such $2$-complex $\Gamma$ we define the quantum configuration space $\mathcal{A}_\Gamma$ to be a direct product of as many copies of $G$ as there are pairs $(e, f)$ such that the edge $e$ is part of the boundary of the face $f$. The set of all these pairs will be denoted as $E\ltimes F$, so we have
\begin{eqnarray}
\mathcal{A}_\Gamma\;:=\;G^{E\ltimes F}.
\end{eqnarray}

\noindent Note that this results in not one link variable, representing the parallel transport as an element of the gauge group $G$, along the edge $e$, but in as many link variables $h_{ef}$ as there are faces meeting at $e$. For the special cases of BF or gauge theories these will all set to be equal. For the general case, the way that these different parallel transports are allowed to depend on the face is precisely the imposition of the simplicity constraints.

For any pair of two complexes $\Gamma\leq \Gamma'$ one defines the map $\pi_{\Gamma'\Gamma}:\mathcal{A}_{\Gamma'}\to\mathcal{A}_\Gamma$ via
\begin{eqnarray}\label{Eq:CoarseGrainingMap}
\pi_{\Gamma'\Gamma}\big(\{h_{e'f'}\}\big)_{ef}\;:=\;\overleftarrow{\prod_{e'\subset e,\,f'\subset f}}h_{e'f'}^{[e',e]}.
\end{eqnarray}

\noindent Here $[e',e]=\pm1$ is the relative orientation of $e'$ and $e$. The $\pi_{\Gamma'\Gamma}$ constitute a coarse graining procedure, and determine how the (microscopic) degrees of freedom in $\mathcal{A}_{\Gamma'}$ combine to the the (macroscopic) degrees of freedom in $\mathcal{A}_{\Gamma}$. Note how in our convention the order of the product is from the right to the left, as befits parallel transports. Note also how, for $\Gamma\leq\Gamma'\leq\Gamma''$ one has that
\begin{eqnarray}\label{Eq:ProjectionConsistency}
\pi_{\Gamma'\Gamma}\pi_{\Gamma''\Gamma'}\;=\;\pi_{\Gamma''\Gamma}.
\end{eqnarray}

\noindent With this data it is possible to define the projective limit of the family $\{\mathcal{A}_\Gamma,\pi_{\Gamma'\Gamma}\}$, to be
\begin{eqnarray}\label{Eq:DefinitionContinuumLimit}
\overline{\mathcal{A}}\;:=\;\lim_{\Gamma\leftarrow}\mathcal{A}_\Gamma\;:=\;\big\{\{a_\Gamma\}_\Gamma\,\big|
\,a_\Gamma\in\mathcal{A}_\Gamma,\;\pi_{\Gamma'\Gamma}a_{\Gamma'}=a_\Gamma\textrm{ whenever }\Gamma\leq\Gamma'\big\}.
\end{eqnarray}

\noindent The space $\overline{\mathcal{A}}$ represents the quantum continuum connections for the gauge group $G$. Note that the limit does
not constitute of performing any limiting procedure in the analytical sense. Rather, a continuum connection $A:=\{a_\Gamma\}_\Gamma\in\overline{\mathcal{A}}$ is given by the collection of all its partial representatives $a_\Gamma$, which satisfy the consistency conditions $\pi_{\Gamma'\Gamma}a_{\Gamma'}=a_\Gamma$, whenever $\Gamma\leq\Gamma'$. Note that there is an associated projection $\pi_\Gamma:\overline{\mathcal{A}}\to\mathcal{A}_\Gamma$ for each $\Gamma$, given by
\begin{eqnarray}\label{Eq:ProjectionMap}
\pi_\Gamma\big(\{a_{\Gamma'}\}_{\Gamma'}\big)\;:=\;a_\Gamma
\end{eqnarray}

\noindent satisfying $\pi_{\Gamma'\Gamma}\pi_{\Gamma'}=\pi_\Gamma$. See appendix \ref{App:Properties} for more information about $\overline{\mathcal{A}}$.

\paragraph*{Path integrals on $\overline{\mathcal{A}}$:}

There is a well-understood way to construct (regular Borel) measures $\mu$ on $\overline{\mathcal{A}}$, which have the interpretation of path-integral measures in quantum theory, or probability measures in statistical physics.

First we note that, given a regular Borel measure $\mu$ on $\overline{\mathcal{A}}$, then we get a family of measures $\{\mu_\Gamma\}$ on the $\mathcal{A}_\Gamma$ via push-forward with the map (\ref{Eq:ProjectionMap}), i.e.
\begin{eqnarray}
\mu_{\Gamma}:=(\pi_\Gamma)_*\mu.
\end{eqnarray}

\noindent Clearly, these measures satisfy the so-called cylindrical consistency condition
\begin{eqnarray}\label{Eq:CylindricalConsistency}
(\pi_{\Gamma'\Gamma})_*\mu_{\Gamma'}\;=\;\mu_\Gamma\quad\textrm{whenever }\Gamma\leq\Gamma'.
\end{eqnarray}

\noindent A collection of measures $\{\mu_\Gamma\}_\Gamma$ satisfying (\ref{Eq:CylindricalConsistency}) is also called a promeasure. We are in the lucky situation that the converse is also true: given a promeasure, then this comes from a unique measure $\mu$ on $\overline{\mathcal{A}}$. See appendix \ref{App:Properties} for details.\\[5pt]

\paragraph*{Observables:} In this context, it is straightforward to define what observables should be: they should correspond to continuous functions on $\overline{\mathcal{A}}$. The sup-norm provides a $C^*$-topology for that space, and a dense subset is given by \emph{continuous cylindrical functions}. These are complex-valued functions $\mathcal{O}$ on $\overline{\mathcal{A}}$ such that there is a $2$-complex $\Gamma$ and a continuous function $\mathcal{O}_\Gamma\in C^0(\mathcal{A}_\Gamma)$ such that
\begin{eqnarray}\label{Eq:CylindricalObervable}
\mathcal{O}\;=\;\mathcal{O}_\Gamma\,\pi_\Gamma.
\end{eqnarray}

\noindent Such an observable is also called \emph{cylindrical over $\Gamma$}. The expectation values of cylindrical functions are then the mathematical realizations of observables in the path integral
\begin{eqnarray}\label{Eq:ExpectationValueObservable}
\langle\mathcal{O}\rangle\;=\;\int_{\overline{\mathcal{A}}}d\mu\;\mathcal{O}\;=\;\int_{\mathcal{A}_\Gamma}d\mu_\Gamma\;\mathcal{O}_\Gamma
\;=:\;\langle\mathcal{O}_\Gamma\rangle_\Gamma.
\end{eqnarray}

It should be noted that, whenever $\mathcal{O}$ is cylindrical over $\Gamma$, then it is automatically cylindrical over all $\Gamma'\geq \Gamma$, because of (\ref{Eq:ProjectionConsistency}):
\begin{eqnarray}
\mathcal{O}\;=\;\mathcal{O}_\Gamma\,\pi_\Gamma\;=\;
\underbrace{\big(\mathcal{O}_\Gamma\,\pi_{\Gamma'\Gamma}\big)}_
{=\,\mathcal{O}_{\Gamma'}}\,\pi_{\Gamma'}.
\end{eqnarray}

\noindent This fact will play a crucial role later, when we consider diffeomorphism-invariant measures $\mu$.

\paragraph*{Physical dictionary and renormalization:}

In this section we would like to recap the physical interpretation of the mathematical concepts introduced in the last section. In particular we would like to demonstrate how they form the natural framework for renormalization in a background-independent context, similarly to the concepts described in \cite{Oeckl:2002ia, Manrique:2005nn, Rovelli:2010qx}.

The space of quantum continuum connections $\overline{\mathcal{A}}$ contains the full degrees of freedom of the theory, and thusly is quite large -- in particular it is no (not even an infinite-dimensional) manifold. However, each $2$-complex $\Gamma$ provides a cut-off, in the sense that is specifies a finite amount of degrees of freedom of the theory, the $h_{ef}$ with $e\subset f\subset \Gamma$. The corresponding configuration space $\mathcal{A}_\Gamma$ is a finite-dimensional manifold, and hence much more accessible than $\overline{\mathcal{A}}$. It can be regarded as a ``finite-dimensional slice through $\overline{\mathcal{A}}$''. The ordering relation $\Gamma\leq\Gamma'$ provides a hierarchy between the degrees of freedom. In particular, the coarse graining maps $\pi_{\Gamma'\Gamma}$ provide the information on how microscopic degrees of freedom, associated to the ``finer'' $2$-complex $\Gamma'$,  combine to macroscopic ones, which are associated to the ``coarser'' $2$-complex $\Gamma$.

In this sense the $2$-complexes $\Gamma$ are the natural analogue of the notion of ``scale'' in the background-independent context.\\[5pt]

As we have already stated, the measure $\mu$ provides the full path integral measure for the continuum. On the other side, the $\mu_\Gamma$ correspond to the effective path integral measures for the degrees of freedom in $\mathcal{A}_\Gamma$. In this sense, the continuum theory is given by the collection of all effective theories at all scales. The cylindrical consistency conditions (\ref{Eq:CylindricalConsistency}) then give a relation between the effective integration measures on $\Gamma$ and $\Gamma'$, providing a background-independent version of Wilson's renormalization group flow.

To see this in a familiar context, consider standard lattice gauge theory as an example. Given, say, a regular $n$-torus with a background being a flat metric having volume $L^n$, consider the set of all regular hypercubic lattices, centered at a common point, having lattice length $a= 2^{-k}N$ for $k\in\mathbb{N}$. To the lattice with lattice length $a$ associate the $2$-complex $\Gamma_a$ consisting of its edges and faces. Consider theories of the form
\begin{eqnarray}
d\mu_{\Gamma_a}\;=\;\left(\prod_{e\subset f}dh_{ef}\right)\left(\int\prod_edU_e\right)\left(\prod_{e\subset f}\delta(h_{ef},U_e)\right)\prod_e\exp\left(-S^{\vec g(a)}(U_e)\right).
\end{eqnarray}

\noindent The function $S^{\vec g(a)}$ is called an \emph{action}, and it is an $a$-dependent function of the link variables $U_e$. More specifically, it can depend on several parameters $\vec g=(g_1,\ldots,g_N)$, called \emph{coupling constants}  which in turn are  $a$-dependent. Condition (\ref{Eq:CylindricalConsistency}) then becomes
\begin{eqnarray}
\exp\left(-S^{\vec{g}(a)}(U_e)\right)\;=\;\int\left(\prod_{e'} dU_{e'}\right)\left(\prod_{e}\delta\left(U_e,\overleftarrow{\prod_{e'\supset e}U_{e'}}\right)\right)\;\exp\left(-S^{\vec{g}(a')}(U_{e'})\right),
\end{eqnarray}

\noindent which can be readily seen to be precisely the RG flow equation for effective lattice actions, in terms of their coupling constants. \\[5pt]

\begin{figure}[hbt!]
\centering
\def\svgscale{0.75}
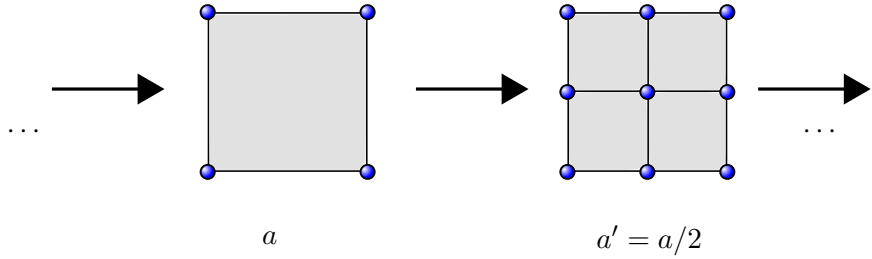
\caption{If there is a background structure, then the RG flow can be organized along a length parameter. }\label{Fig:Figure02}
\end{figure}

%

In the general case, when there is no background metric, one has to keep the collection of all $2$-complexes $\Gamma$, together with the information of how they are embedded into each other. In this case, the condition for there to be a continuum measure $\mu$ is still given by the cylindrical consistency conditions (\ref{Eq:CylindricalConsistency}), but instead of an easy set of equations one now generically has an uncountable set of equations for an uncountable set of measures.

\begin{figure}[hbt!]
\centering
\def\svgscale{0.75}
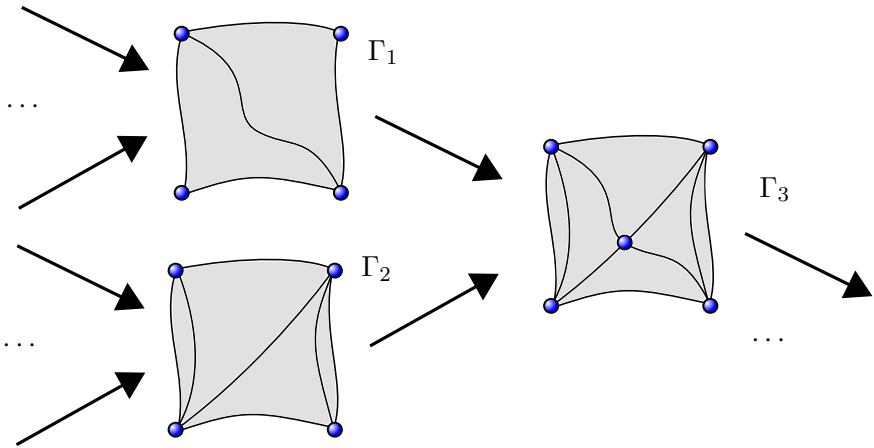
\caption{Without a background structure, the RG flow runs not along a single parameter, but along the partially ordered set of all $2$-complexes embedded in $\mathcal{M}$.}\label{Fig:Figure03}
\end{figure}

%

So, while the RG flow in the background-dependent case runs along a single parameter (the length scale $a$, figure \ref{Fig:Figure02}), in the background-independent context the RG flow runs along the partially ordered set of all $2$-complexes $\Gamma$ (see figure \ref{Fig:Figure03}). Note an important difference to the approach in \cite{Oeckl:2002ia}: The flow described here is not invertible, because during the coarse graining process information about microscopic degrees of freedom is lost. In \cite{Oeckl:2002ia}, the flow is defined to go in both directions, leading to the notion of a renormalization groupoid.

We will see an example for this general case in section \ref{Sec:Example}, and a discussion for how to simplify this complicated collection of equations in section \ref{Sec:Approximations}.

\section{Diffeomorphism invariance}\label{Sec:DiffInvariance}

The group of space-time diffeomorphisms is a gauge symmetry of general relativity. It is natural to demand that it should also appear as a symmetry of the path integral measure. While one can achieve this on a formal level in the linearized regime \cite{Veltman:1975vx}, in general it is not known how to construct a non-perturbative path integral measure with full diffeomorphism symmetry.

The first example of a rigorous diffeomorphism-invariant measure on the space of all connections (i.e.~not just flat ones) is the construction in \cite{Ashtekar:1994mh, Ashtekar:1995zh}, which led to the remarkable success of loop quantum gravity. The measure in question, however, is not the path integral measure for covariant quantum gravity, but rather the kinematical measure for canonical quantum gravity. Consequently, it is invariant under diffeomorphisms of the Cauchy surface only, and contains no knowledge about the dynamics. Still, key elements of its construction play a central role in our approach as well.

In four space-time dimensions, the tension between continuum diffeomorphism symmetry on the one side, and the discretization in some quantum gravity approaches on the other side is deep \cite{Dittrich:2008pw}. Incorrect implementation of diffeomorphism symmetry on the discrete level can lead to the emergence of spurious degrees of freedom, as one would expect from a breaking of classical gauge symmetry in the quantum theory \cite{Bahr:2009ku, Bahr:2011xs}. Still, from lower-dimensional models there are hints that at the IR fixed point diffeomorphism symmetry can emerge \cite{Bahr:2009qc, Bahr:2009mc}, and that one can even construct the discrete theories with full continuum diffeomorphism symmetry, leading to the quantization of discretizations of the correct continuum degrees of freedom \cite{Bahr:2009qc, Bahr:2011uj, Rovelli:2011fk}, although this might be highly nontrivial in four space-time dimensions \cite{Dittrich:2011vz, Dittrich:2012qb}.

It should be noted that succeeding in this would be equivalent to constructing a nontrivial four-dimensional manifold invariant \cite{Barrett:1995mg, Pfeiffer:2004pe}. For these not many nontrivial examples are known, in particular not if the invariant should not also be topological.\\[5pt]

In our framework, it is easy to define what is meant by diffeomorphism-invariant path-integral measures, and in the following we will see that the condition for diffeomorphism-invariance will pose severe restrictions on the partial measures $\mu_\Gamma$. For this, the fact that we have kept the information about how each $2$-complex $\Gamma$ is embedded in the manifold $\mathcal{M}$, is a necessary requirement.

Denote the group of semi-analytic diffeomorphisms by ${\rm Diff}^{\omega/2}(\mathcal{M})$, then the action of this group on $\mathcal{M}$ can be pulled back to an action on the space of $2$-complexes. So e.g.~for each edge $e$ in $\Gamma$, there is a corresponding edge $\phi(e)$ in $\phi(\Gamma)$, and so on. In the following we assume that the family of $2$-complexes considered is large enough, so that for each $\Gamma$ and each $\phi\in{\rm Diff}^{\omega/2}(\mathcal{M})$, the $2$-complex $\phi(\Gamma)$ is also a member of that family. If that is the case, $\phi\in{\rm Diff}^{\omega/2}(\mathcal{M})$ induces a Lie group isomorphism between $\mathcal{A}_\Gamma$ and $\mathcal{A}_{\phi(\Gamma)}$. It is not hard to see that this map respects the ordering relation (\ref{Eq:OrderingRelation}), i.e.~if $\Gamma\leq\Gamma'$, then $\phi(\Gamma)\leq\phi(\Gamma')$.

%

\begin{figure}[hbt!]
\centering
\def\svgscale{0.5}
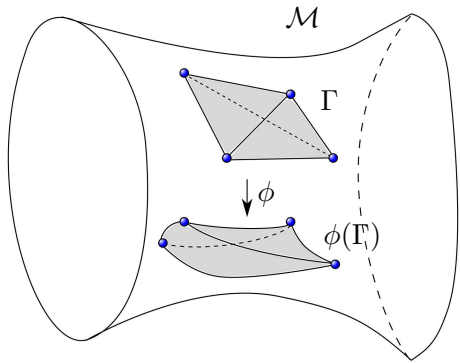
\caption{Diffeomorphisms $\phi$ of $\mathcal{M}$ act naturally on $\overline{\mathcal{A}}$ via the $2$-complexes $\Gamma$.}\label{Fig:Figure04}
\end{figure}

Consequently, the action of $\phi$ can be pulled back to  $\overline{\mathcal{A}}$: If $A\in\overline{\mathcal{A}}$ is given by (\ref{Eq:DefinitionContinuumLimit}), then
\begin{eqnarray}
\phi^*A\;:=\;\{a_{\phi(\Gamma)}\}_\Gamma.
\end{eqnarray}

\noindent Furthermore, given a collection of cylindrically consistent measures $\{\mu_\Gamma\}_\Gamma$, the measures $\{\phi_*\mu_\Gamma\}_{\phi(\Gamma)}$ are also cylindrically consistent, and hence define a measure on $\overline{\mathcal{A}}$. Of course, this measure coincides with the push-forward $\phi_*\mu$, given by the action $\phi^*$ on $\overline{\mathcal{A}}$\\[5pt]

In this context it is a natural question to ask whether a measure $\mu$ is invariant under the action of ${\rm Diff}^{\omega/2}(\mathcal{M})$, i.e.~whether or not the equation $\phi_*\mu=\mu$ holds. For this, take some observable $\mathcal{O}$ cylindrical over a $2$-complex $\Gamma$, i.e.~$\mathcal{O}=\mathcal{O}_\Gamma\,\pi_\gamma$ with $\mathcal{O}_\Gamma\in C^0(\mathcal{A}_\Gamma)$. Also, for some semi-analytic diffeomorphism $\phi$, there exists, by definition, a $2$-complex $\Gamma'$ which is finer than both $\Gamma$ and $\phi(\Gamma)$, i.e. $\Gamma,\phi(\Gamma)\leq\Gamma'$. Then the two observables
\begin{eqnarray*}
\mathcal{O}_1\;&:=&\;\mathcal{O}_\Gamma\,\pi_{\Gamma'\Gamma}\\[5pt]
\mathcal{O}_2\;&:=&\;\big(\phi_*\mathcal{O}_\Gamma\big)\,\pi_{\Gamma'\phi(\Gamma)}
\end{eqnarray*}
\noindent are both cylindrical over $\Gamma'$, and from $\phi_*\mu=\mu$ and (\ref{Eq:ExpectationValueObservable}) it is easy to deduce that they should have the same expectation value, i.e.~
\begin{eqnarray}
\langle\mathcal{O}_1\rangle_{\Gamma'}\;=\;\langle\mathcal{O}_2\rangle_{\Gamma'}.
\end{eqnarray}

\noindent It is important to note that this is a strong restriction on $\mu_{\Gamma'}$. In particular, the observables supported on two sub-complexes of $\Gamma'$ which can be connected by a semi-analytic diffeomorphism on $\mathcal{M}$, need to have the same expectation values, when evaluated with the measure $\Gamma'$. For this it is \emph{not} necessary that the whole diffeomorphism maps $\Gamma'$ onto itself, only that it maps the two sub-complexes $\Gamma$ and $\phi(\Gamma)$ onto each other.

\begin{figure}[hbt!]
\centering
\def\svgscale{0.75}
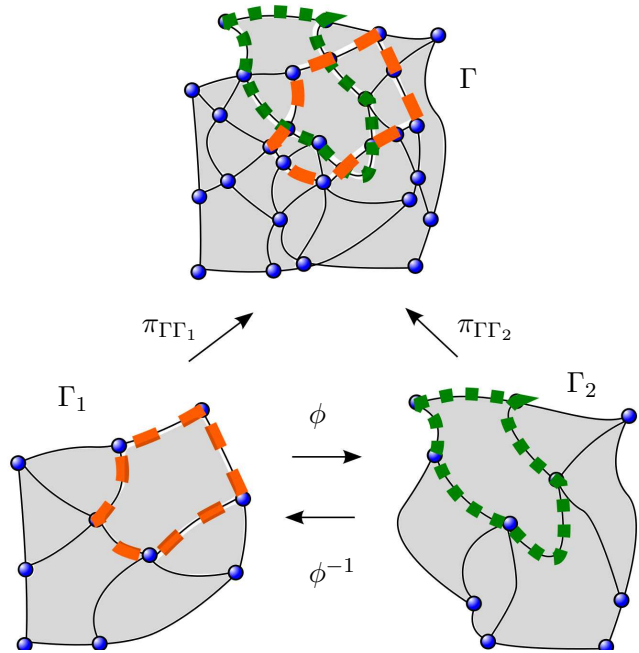
\caption{Both $\Gamma_1$ and $\Gamma_2=\phi(\Gamma_1)$ can be refined by $\Gamma$ due to cylindrical consistency. Diffeomorphism invariance of expectation values of observables then translates into strong conditions for the measure $\mu_{\Gamma}$.
   }\label{Fig:Figure05}
\end{figure}
%
%

The converse is also true: Let $\mu$ be a measure on $\overline{\mathcal{A}}$ with the following property: For each $\Gamma$, and each $\Gamma_1,\,\Gamma_2\leq\Gamma$ such that there exists $\phi\in\textrm{Diff}^{\omega/2}(\mathcal{M})$ such that $\phi(\Gamma_1)=\Gamma_2$, one has that
\begin{eqnarray}\label{Eq:ConditionDiffInvariance}
\langle\mathcal{O}_1\,\pi_{\Gamma_1\Gamma}\rangle_\Gamma\;=\;\langle\mathcal{O}_2\,\pi_{\Gamma_2\Gamma}\rangle_\Gamma
\end{eqnarray}

\noindent for each $\mathcal{O}_1$ cylindrical over $\Gamma_1$, and $\mathcal{O}_2:=\phi(\mathcal{O}_1)$. Then $\phi_*\mu=\mu$ for all $\phi\in \textrm{Diff}^{\omega/2}(\mathcal{M})$.

In other words, diffeomorphism-invariance can be checked separately for each $\mu_\Gamma$. It should be noted, however, that in order to do so, it is not sufficient to know just the combinatorial data of $\Gamma$. One also needs the information about how $\Gamma$ embeds into $\mathcal{M}$.

\section{Examples}\label{Sec:Example}


In the following we consider a very simple example, which can be treated completely and analytically. For more interesting systems we do not expect that the RG equations can be solved in full generality.

Let $\mathcal{M}$ be a $2$-dimensional, connected, closed, oriented, semi-analytic manifold. The family of $2$-complexes we consider consists of all semi-analytic cellular decompositions of $\mathcal{M}$, i.e.~ $2$-complexes that completely ``fill out'' $\mathcal{M}$. We want to consider $U(1)$ gauge theory on (a trivial line bundle over) $\mathcal{M}$, so to a $2$-complex $\Gamma$ with $E$ edges we associate the partial configuration space
\begin{eqnarray*}
\mathcal{A}_\Gamma\;=\;U(1)^E.
\end{eqnarray*}

\noindent In this case, the coarse-graining maps $\pi_{\Gamma'\Gamma}: U(1)^{E'}\to U(1)^E$ are given by
\begin{eqnarray}
\Big(\pi_{\Gamma'\Gamma}(u_1,\ldots, u_{E'})\Big)_e\;=\;\prod_{e'\subset e}u_{e'}^{[e,e']}
\end{eqnarray}

\noindent where $[e,e']=\pm 1$, depending on whether the orientations of $e'\subset e$ agree or disagree. We formulate the RG equations considering the family of measures
\begin{eqnarray}\label{Eq:ExampleMeasure}
d\mu_{\Gamma}^{\vec a,\vec\theta}\;:=\;\prod_{f\in\Gamma^{(2)}}K_{a_f}\big(H_fe^{i\theta_f}\big)\;du_1\ldots,du_E
\end{eqnarray}

\noindent where the product ranges over all $F$ faces of $\Gamma$, $a_1,\ldots, a_F>0$ and $\theta_1,\ldots,\theta_F\in\mathbb{R}$. As customary, we denote by $K_t:U(1)\to[0,\infty)$ the heat kernel (on $U(1)$)
\begin{eqnarray}\label{Eq:HeatKernelU(1)}
K_t(u)\;=\;\sum_{n\in\mathbb{Z}}e^{-n^2\frac{t}{2}}\,u^n
\end{eqnarray}

\noindent at time $t>0$, and by $H_f$ the ordered product of $u_e$ of edges $e$ around the face $f$.

The observables in this case are simply given by continuous functions on $\overline{\mathcal{A}}$, a  dense set of which is generated by the so-called \emph{charge-network-functions} on a graph $\gamma=\Gamma^{(1)}$.
\begin{eqnarray}\label{Eq:ExampleObservable}
\mathcal{O}^{\vec{n}}(u_1,\ldots,u_E)\;=\;\prod_e u_e ^{n_e}
\end{eqnarray}

\noindent with $n_1,\ldots,n_E\in\mathbb{Z}$. It is an elementary calculation to show that the expectation value of (\ref{Eq:ExampleObservable}) with the measure (\ref{Eq:ExampleMeasure}) is given by
\begin{eqnarray}\label{Eq:ExampleExpectationValue}
\langle\mathcal{O}^{\vec n}\rangle_{\vec a,\vec \theta}\;=\;\sum_{n_f}\Big(\prod_f e^{-{n_f^2\frac{a_f}{2}}}\,e^{in_f\theta_f}\Big)\prod_e\delta_{n_e+[e,f_1]n_{f_1}+[e,f_2]n_{f_2},0}
\end{eqnarray}

\noindent where for an edge $e$ we denote by $f_{1,2}$ the two faces that meet at $e$. Using the orientability of $\mathcal{M}$, one can see that (\ref{Eq:ExampleExpectationValue}) vanishes if $\mathcal{O}^{\vec{n}}$ is not gauge-invariant, i.e.~if at some vertex $v$ the condition
\begin{eqnarray}\label{Eq:EyampleGaugeInvariance}
\sum_{e\supset v}n_e^{[v,e]}\;\stackrel{!}{=}\;0
\end{eqnarray}
\noindent is violated. If (\ref{Eq:EyampleGaugeInvariance}) holds, on the other hand, then (\ref{Eq:ExampleExpectationValue}) can be evaluated by computing just one sum over $\mathbb{Z}$.

Using an elementary property of the heat kernel, it is easy to show that the RG equations
\begin{eqnarray}
\langle\mathcal{O}^{\vec n}\rangle_{\vec a,\vec \theta}\;=\;\langle\mathcal{O}^{\vec n}\,\pi_{\Gamma'\Gamma}\rangle_{\vec a',\vec \theta'}\qquad\text{for all }\vec n
\end{eqnarray}
for a pair $\Gamma\leq\Gamma'$ of $2$-complex are given by
\begin{eqnarray}
\begin{aligned}
a_{f}\;=&\;\sum_{f'\subset f}a_{f'}\\[5pt]
\theta_f\;=&\;\sum_{f'\subset f}[f',f]\theta_{f'}
\end{aligned}\label{Eq:ExampleRGEquations}
\end{eqnarray}

\noindent where $[f',f]=\pm 1$, depending on whether the orientations of $f'\subset f$ agree or disagree.\\[5pt]

There is quite a simple set of solutions to (\ref{Eq:ExampleRGEquations}): Choose a semi-analytic Riemannian metric $g\in {\rm Sym}^2T^*\mathcal{M}$ and a semi-analytic $2$-form $\theta\in \Omega^2(\mathcal{M})$,\footnote{These are, of course, not the only solutions: One can relax or strengthen the semi-analyticity conditions and choose e.g. smooth or continuous functions instead.} and define
\begin{eqnarray}
\begin{aligned}
a_f\;&:=\;\int_f\sqrt{\det g}\\[5pt]
\theta_f\;&:=\;\int_f\theta
\end{aligned}\label{Eq:ExampleRGSolution}
\end{eqnarray}

\noindent It is straightforward to see that (\ref{Eq:ExampleRGSolution}) satisfy (\ref{Eq:ExampleRGEquations}), defining a measure $\mu^{g,\theta}$ on $\overline{\mathcal{A}}$.

There are two interesting limiting cases of the solutions (\ref{Eq:ExampleRGSolution}). One corresponds to formally taking $g\to \infty$, i.e.~all $a_f\to \infty$, while keeping $\theta$ fixed. The result exists as a measure on $\overline{\mathcal{A}}$ and can be easily identified with the Ashtekar-Lewandowski measure $\mu_{\rm AL}$ \cite{Ashtekar:1994mh, Ashtekar:1995zh}, also called \emph{high temperature fixed point}.

Another limiting solution is comprised of the limit $g\to 0$, i.e.~of taking all $a_f\to 0$. It can be shown that the limit measure $\mu^{0,\theta}$ exists as a measure on $\overline{\mathcal{A}}$. In this limit we arrive at a rigorous version of the expression
\begin{eqnarray}\label{Eq:ExampleLimitMeasure}
\langle\mathcal{O}\rangle\;=\;\frac{1}{Z}\int_{\overline{\mathcal{A}}}\mathcal{D}A\;\delta\big(F[A]-\theta\big)\;
\mathcal{O}[A].
\end{eqnarray}

\noindent In the case of $\theta=0$, this becomes simply the measure of $U(1)$ BF theory on $\mathcal{M}$. In the case of $G$ being finite instead of $G=U(1)$, and $\theta=0$, the state sum $Z$ in (\ref{Eq:ExampleLimitMeasure}) is finite and becomes the well-known Dijkgraaf-Witten invariant of $2$-dimensional manifolds \cite{Dijkgraaf:1989pz}.

\paragraph*{Diffeomorphism-invariance:}

It can be easily shown that for a semi-analytic diffeomoprhism $\phi$ one has that
\begin{eqnarray}
\phi_*\mu^{g,\theta}\;=\;\mu^{\phi^*g,\phi^*\theta}
\end{eqnarray}

\noindent so generically the continuum measures $\mu^{g,\theta}$ on $\overline{\mathcal{A}}$ are not diffeomorphism-invariant.

\begin{figure}[hbt!]
\centering
\def\svgscale{0.5}
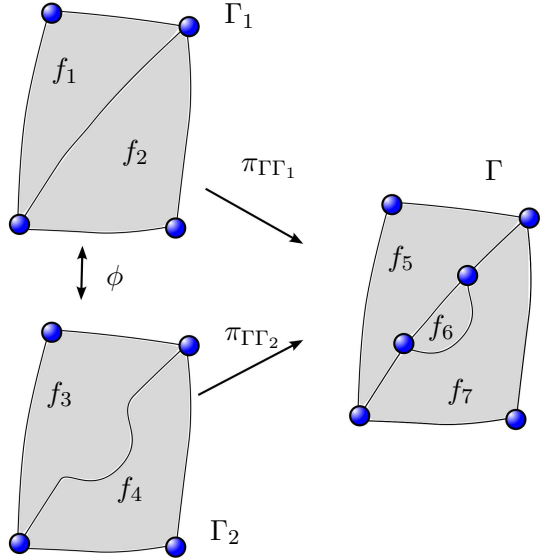
\caption{Diffeomorphism-invariance and cylindrical consistency lead to $a_{f_6}=0$ or (formally) $a_{f_6}=\infty$.
    }\label{Fig:Figure06}
\end{figure}

In particular, the condition for diff-invariance imposes further restrictions, additional to the RG flow equations (\ref{Eq:ExampleRGEquations}). To see this, we consider a $2$-complex $\Gamma_1$ and a diffeomorphism $\phi$ which is different from the identity map only in a small neighborhood of an interior point of an edge $e$, moving the edge slightly (see figure \ref{Fig:Figure06}). Denote by $\Gamma$ the obvious $2$-complex finer than $\Gamma_1$ and $\Gamma_2:=\phi(\Gamma_1)$, and consider a charge-network function $\mathcal{O}^{\vec{n}}$ on $\Gamma_1$.

%
%

From figure \ref{Fig:Figure06} it can be seen that the condition
\begin{eqnarray}
\langle\mathcal{O}^{\vec{n}}\,\pi_{\Gamma\Gamma_1}\rangle_{\Gamma}
\;=\;\langle(\phi^*\mathcal{O}^{\vec{n}})\pi_{\Gamma\Gamma_2}\rangle_{\Gamma}
\end{eqnarray}

\noindent for all $\vec n\in\mathbb{Z}^E$ leads to
\begin{eqnarray*}
\begin{aligned}
a_{f_1}\;&=\;a_{f_3} & a_{f_2}\;&=\;a_{f_4}\\[5pt]
a_{f_1}\;&=\;a_{f_5} & a_{f_2}\;&=\;a_{f_6}+a_{f_7}\\[5pt]
a_{f_3}\;&=\;a_{f_5} + a_{f_6}& a_{f_4}\;&=\;a_{f_7}
\end{aligned}
\end{eqnarray*}

\noindent and similar equations for the $\theta_f$. Since these equations have to hold for every edge, relations for all $a_f$ (and all $\theta_f$) for all $\Gamma$ can be deduced. It's not hard to see that the only solutions for this are either (formally) $a_f=\infty$ for all $f$, or, if $a_f$ is finite, then $a_{f_6}=-a_{f_6}$, hence
\begin{eqnarray*}
a_f=0\qquad\text{for all }f.
\end{eqnarray*}

\noindent These two solutions correspond (formally) to the measures $\mu_{\rm AL}$ and $\mu^{0,0}$. Indeed, it is not difficult to check that both the Ahtekar-Lewandowski measure $\mu_{\rm AL}$ and the BF-theory measure $\mu^{0,0}$ are invariant under diffeomoprhisms.

We see that imposing diffeomorphism-invariance has drastically reduced the allowed solutions, by introducing new equations, additionally to the RG flow equations. \\[5pt]

\section{Approximation schemes}\label{Sec:Approximations}

We have seen that there are uncountably many RG flow equations given by the cylindrical consistency conditions (\ref{Eq:CylindricalConsistency}). In the easy example from section \ref{Sec:Example} it was possible to write them in a closed form for a specific set of measures (\ref{Eq:ExampleMeasure}), and to guess some very obvious solutions. This was due to the easy nature of the heat kernel on $U(1)$ in 2d, and will certainly not be possible in more interesting scenarios.

As in almost all cases of physical theories, one needs to make approximations to be able to perform -- even numerical -- calculations. In the following we will present three different ways of approximating the RG flow equations (\ref{Eq:CylindricalConsistency}), making them more accessible to analytical and numerical investigation. Afterwards we will apply all of them in order to treat a spin foam model with quartic interaction term.

\paragraph*{1.) Restricting the $2$-complexes}

One reason for the complexity of the full set of the RG equations is that the set of $2$-complexes is not only uncountable, they can also be completely irregular, making bookkeeping difficult. So instead of taking on all $2$-complexes at the same time, one could restrict the set under consideration, simplifying the analysis immensely.

One example would certainly be considering a sequence of regular lattices which are sublattices of each other, as in the regular hypercubic calculations in lattice gauge theory. The difference is that there would be no lattice length attached to it, rather, the lattice length would be one observable among many. In this case it might be possible to retrieve standard RG flow equations in the sense of ``evolving constants'' which lie at the heart of the relational framework for Dirac observables \cite{Rovelli:2001bz, Dittrich:2005kc}.

Although a severe restriction will almost certainly violate the condition at the beginning of section \ref{Sec:DiffInvariance}, which was necessary to define the action of the diffeomorphisms, one can still look for diffeomorphism-invariant measures in this case, because the condition for diffeo-invariance can be pulled back to each single $2$-complex $\Gamma$, even e.g.~regular lattices, via (\ref{Eq:ConditionDiffInvariance}).

\paragraph*{2.) Restricting the observables}

Given $\Gamma\leq\Gamma'$, then the condition $(\pi_{\Gamma'\Gamma})_*\mu_{\Gamma'}=\mu_\Gamma$ can be rephrased as
\begin{eqnarray}\label{Eq:CylindricalConsistencyAlternative}
\langle\mathcal{O}_\Gamma\,\pi_{\Gamma'\Gamma}\rangle_{\Gamma'}\;=\;\langle \mathcal{O}_\Gamma\rangle_\Gamma\qquad\textrm{for all }\mathcal{O}_\Gamma\in C^0(\mathcal{A}_\Gamma).
\end{eqnarray}

\noindent Instead of demanding (\ref{Eq:CylindricalConsistencyAlternative}) to hold for all observables $\mathcal{O}_\Gamma$, one can select a small subset of them, and require (\ref{Eq:CylindricalConsistencyAlternative}) to hold only for those. By considering only a few important observables, the problem can be simplified considerably, as we will see in an example later.

\paragraph*{3.) Restricting the parameter space}

One of the most important approximations, is restricting the space of measures $\{\mu_\Gamma\}_\Gamma$ one formulates equation (\ref{Eq:CylindricalConsistency}) in. This can be done by making an ansatz for the measure, e.g.~in terms of an action function
\begin{eqnarray}\label{Eq:MeasureAnsatz}
d\mu_\Gamma^{\vec g}\;=\;\left(\prod_{e\subset f}dh_{ef}\right)\exp\left(-S^{(\vec{g})}(h_{ef})\right).
\end{eqnarray}

\noindent An example for this would be e.g.~the ansatz (\ref{Eq:ExampleMeasure}) we were using in the example. Even more dramatically would be the restriction to actually finitely many parameters, so that the RG flow equations would be equations for a finite set of coupling constants depending on $\Gamma$, i.e.~$\vec{g}(\Gamma)$. The standard RG equations for lattice gauge theory certainly fall into this category.

In the example in section \ref{Sec:Example} we were able to solve the RG equations (\ref{Eq:CylindricalConsistency}) within this ansatz. In most interesting cases this will not be possible. One of the reasons for that is that any non-topological exact solution to the RG flow equations will necessarily generate non-local couplings \cite{Migdal:1975zg, Kadanoff:1976jb, Dittrich:2014rha}, with the 2d heat kernel being one of the very few exceptions to this rule. In particular, any ansatz which factorizes over the constituents of the $2$-complexes, as is the case with e.g.~all present spin foam models, will not be an exact solution to the RG flow equations, because it will not include non-local couplings.

However, one can still look for approximate solutions in a realm in which non-local couplings are small, and restrict the flow to the space of measures parametrized as (\ref{Eq:MeasureAnsatz}), by demanding that the error
\begin{eqnarray}\label{Eq:ApproximateRGFlowEquation}
\Xi_{\vec g,\vec{g}'}\;:=\;\sup_{\|\mathcal{O}_\Gamma\|=1}\big|\langle\mathcal{O}_\Gamma\rangle_{\vec{g}}
\;-\;\langle\mathcal{O}_\Gamma\,\pi_{\Gamma'\Gamma}\rangle_{\vec{g}'}|\;\stackrel{!}{=}\;\textrm{min}
\end{eqnarray}

\noindent is small, with
\begin{eqnarray}
\langle\mathcal{O}_\Gamma\rangle_{\vec{g}}\;:=\;\int_{\mathcal{A}_{\Gamma}} d\mu_{\Gamma}^{\vec{g}}\;\mathcal{O}_\Gamma,\qquad
\langle\mathcal{O}_\Gamma\,\pi_{\Gamma'\Gamma}\rangle_{\vec{g}'}\;:=\;\int_{\mathcal{A}_{\Gamma'}} d\mu_{\Gamma'}^{\vec{g'}}\;\mathcal{O}_\Gamma\,\pi_{\Gamma'\Gamma}.
\end{eqnarray}
\noindent If one chooses $\|\cdot\|$ to be the supremum norm on $C^0(\mathcal{A}_\Gamma)$, then (\ref{Eq:ApproximateRGFlowEquation}) is just the minimization of the operator norm of the difference of linear functionals  corresponding to the measures $\mu_\Gamma^{\vec g}$ and $(\pi_{\Gamma'\Gamma})_*\mu_{\Gamma'}^{\vec{g}'}$, respectively, via the Riesz representation theorem. By varying this norm one can weigh the errors on some observables to be more severe than on others, introducing more freedom in the procedure.

Note that this approximation is akin to the one employed in the functional renormalization approach to quantum gravity (\cite{Reuter:2012id} and references therein), where the space of allowed actions is kept fixed as well. Of course, this approximation might lead to solutions which are very far away from an actual solution, if the space of measures is restricted too strongly. Just as in the FRG approach, though, one can make a truncation to few parameters, and hope to find nontrivial, i.e.~interacting, UV fixed points of the action. Afterwards, one can enlarge the parameter space and check whether the characteristic fixed point is stable. This has led to quite some success in that approach, so that one can hope that similar procedures can also work for spin foam models.

It should also be noted that, due to the specific nature of this approximation, one can suppress non-local couplings by specifically excluding them from the ansatz for the partial measures $\mu_\Gamma$. This is reminiscent of the Migdal-Kadanoff approximation within lattice gauge theory \cite{Migdal:1975zg, Kadanoff:1976jb}, and the highest Eigenvalue approximation within tensor network renormalization. It would be very interesting to compare these methods with the one presented here, and we hope to come back to this point in future work.

\paragraph*{Example:}

In what follows we will utilize all three of the above approximation methods to compute the RG flow of a simple $2d$ model with gauge group $U(1)$ and quartic interaction term.

Consider a single face $F$ diffeomorphic to the closed disc, bounded by the single edge $E$, forming the $2$-complex $\Gamma$, and two different subdivisions $\Gamma_{1,2}$ of $\Gamma$ into $N_1$ and, respectively, into $N_2>N_1$ faces. For $\mu_{\Gamma_1}$ we make a similar ansatz to (\ref{Eq:ExampleMeasure}), with $a_f\equiv a$ constant for all faces and $\theta_f=0$, including a quartic term
\begin{eqnarray}
d\mu_{\Gamma_1}\;=\;\frac{1}{N}\left(\prod_edg_e\right)\prod_f K_{a_1}(e^{i\phi_f})\,e^{-\lambda_1\,\sin^4\phi_f}
\end{eqnarray}

\noindent with the heat kernel $K_a(u)$ (\ref{Eq:HeatKernelU(1)}), $N$ normalizing the measure to $1$, and
\begin{eqnarray}
g_e\;=\;e^{i\phi_e},\qquad \phi_f:=\sum_{e\subset f}[e, f]\,\phi_e,
\end{eqnarray}

\noindent with all $\phi_e\in [0,2\pi)$. Thus we have two coupling constants, $a_1$ and $\lambda_1$, on $\Gamma_1$. A similar measure can be defined on $\Gamma_2$. We relate the two sets of coupling constants by comparing the expectation values of the observables on $\Gamma$. There is one gauge-invariant observable for each $n\in\mathbb{Z}$ on $\Gamma$, given by
\begin{eqnarray}
\mathcal{O}^n(h_{EF})\;:=\;h_{EF}^n.
\end{eqnarray}

\noindent A straightforward calculation reveals that
\begin{eqnarray}\label{Eq:ApproximateEValue}
\langle \mathcal{O}^n\,\pi_{\Gamma_1\Gamma}\rangle_{\Gamma_1}\;=\;\left(\frac{1}{N}\sum_{k+l=n}e^{-k^2\frac{a_1}{2}}\int_0^{2\pi}
\frac{d\phi}{2\pi}e^{il\phi-\lambda_1 \sin^4\phi}\right)^{N_1}.
\end{eqnarray}

\noindent It is not hard to show that (\ref{Eq:ApproximateEValue}) is symmetric under $n\to-n$, so we only consider $n>0$.

In order to explore the UV limit, we consider large $N_1$ and $N_2$, with $N_2/N_1 \gtrsim 1$, and demand that the difference of expectation values of $\mathcal{O}^n$ are minimal, i.e.
\begin{eqnarray}\label{Eq.ApproximateCondition}
\Xi\;:=\;\sum_{n=1}^\infty c_n\big|\langle \mathcal{O}^n\,\pi_{\Gamma_1\Gamma}\rangle_{\Gamma_1}\;-
\langle \mathcal{O}^n\,\pi_{\Gamma_2\Gamma}\rangle_{\Gamma_2}\big|\;\stackrel{!}{=}\;\textrm{min},
\end{eqnarray}

\noindent with $c_n>0$. Given $a_1,\, \lambda_1$, condition (\ref{Eq.ApproximateCondition}) determines $a_2,\,\lambda_2$.

\begin{figure}[hbt!]
\centering
\def\svgscale{1}
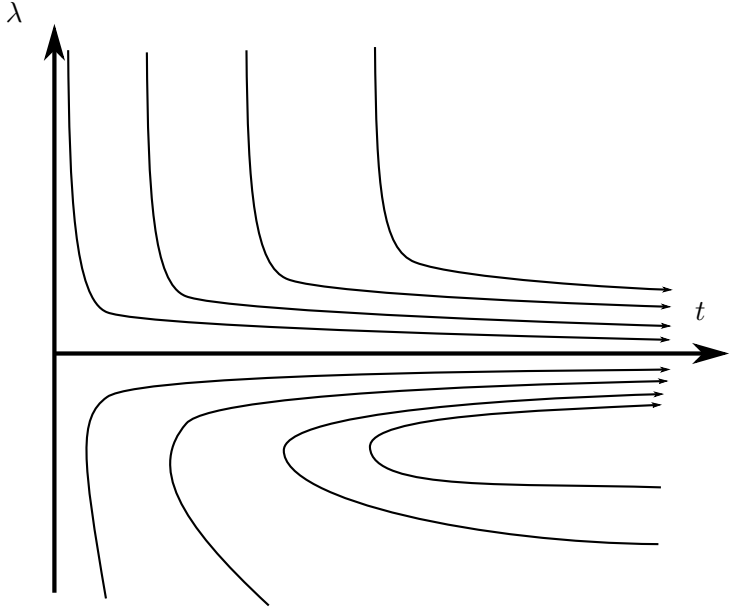
\caption{Qualitative flow of the quartic spin foam model in $d=2$.
    }\label{Fig:Figure07}
\end{figure}

%

\begin{figure}[hbt!]
 \begin{minipage}[t]{0.45\linewidth}
\includegraphics[scale=0.75]{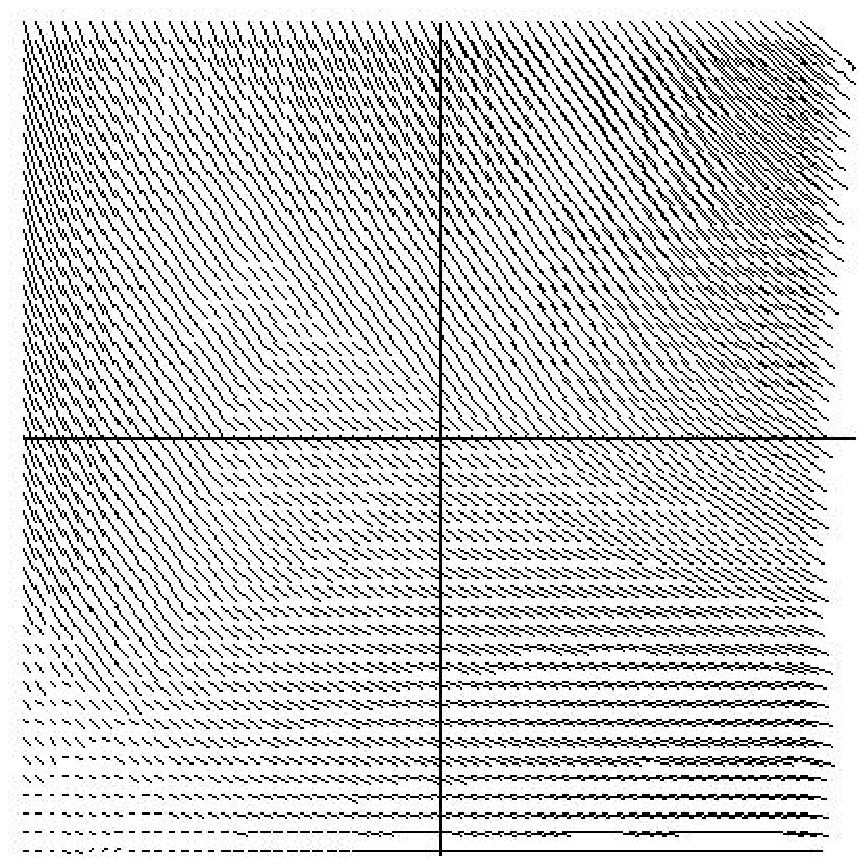}
\caption{Numerical flow for $c_1=c_2=1$ in the range of $1\leq t\leq 4$, $0\leq\lambda\leq 2.25$.}
\label{Fig:Figure08}

 \end{minipage}\hspace{15pt}
 \begin{minipage}[t]{0.45\linewidth}

\includegraphics[scale=0.75]{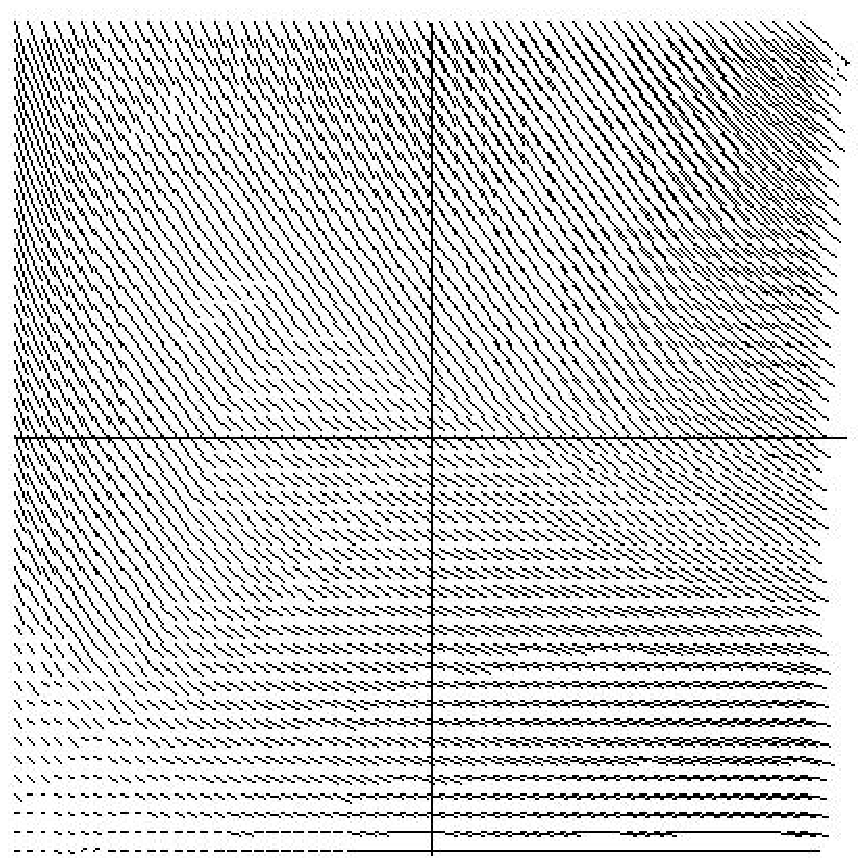}
\caption{Numerical flow for $c_1=c_2=c_3=1$ in the range of $1\leq t\leq 4$, $0\leq\lambda\leq 2.25$.}
\label{Fig:Figure09}
 \end{minipage}
 \end{figure}

\begin{figure}[hbt!]
 \begin{minipage}[t]{0.45\linewidth}
\includegraphics[scale=0.75]{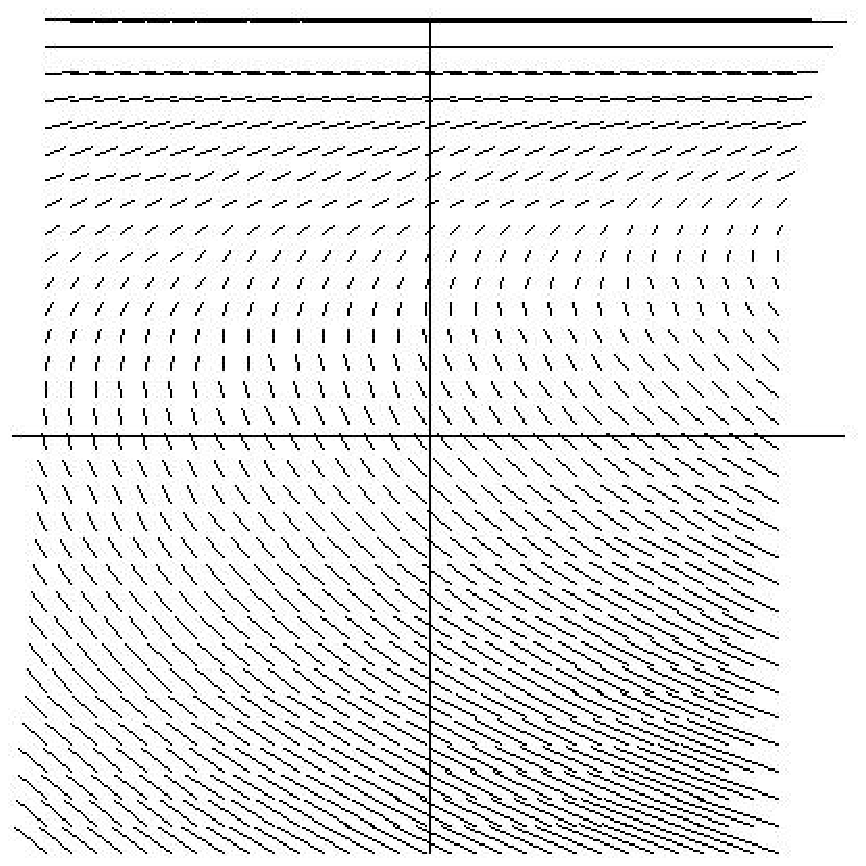}
\caption{Numerical flow for $c_1=c_2=1$ in the range of $4\leq t\leq 5.5$, $-1.5\leq\lambda\leq 0.1$.}
\label{Fig:Figure10}
 \end{minipage}\hspace{15pt}
 \begin{minipage}[t]{0.45\linewidth}

\includegraphics[scale=0.75]{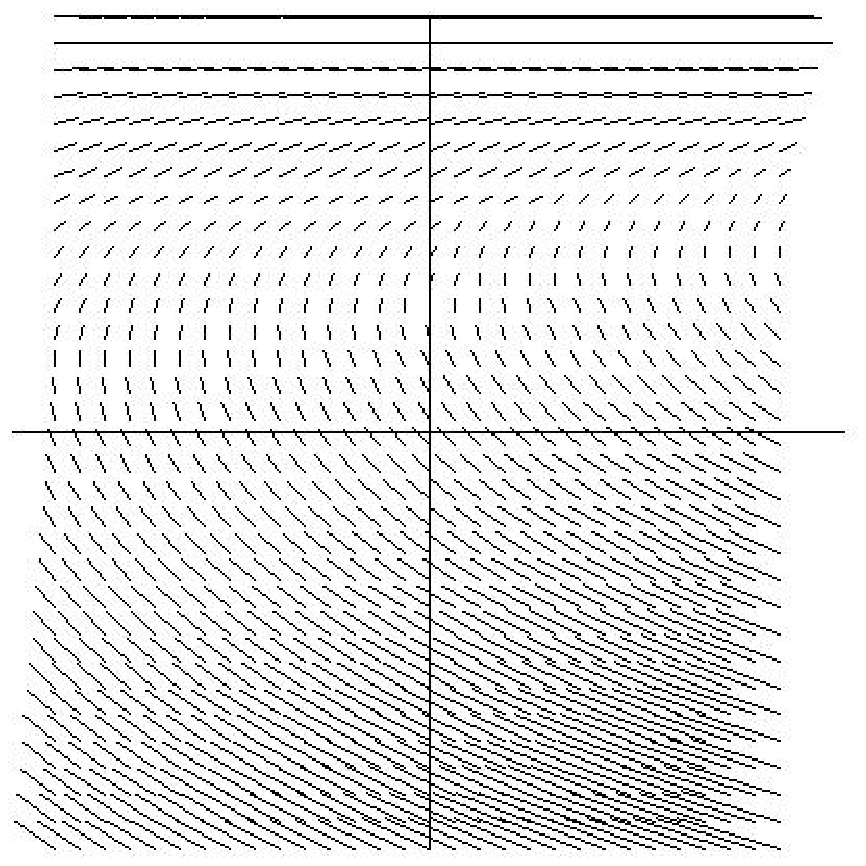}
\caption{Numerical flow for $c_1=c_2=1, c_3=4$ in the range of $4\leq t\leq 5.5$, $-1.5\leq\lambda\leq 0.1$.}
\label{Fig:Figure11}
 \end{minipage}
 \end{figure}

Within a numerical treatment, we considered several cases of $c_n$, finding the RG flows being relatively stable under change of the $c_n$. A depiction of the phase diagram is shown in figure \ref{Fig:Figure07}, while in figures \ref{Fig:Figure08}, \ref{Fig:Figure09}, \ref{Fig:Figure10}, \ref{Fig:Figure11} some numerical flows with $N_2/N_1=1.02$ are presented. Note that, due to the compactness of $U(1)$, the quartic coupling $\lambda$ can be negative.\\[5pt]

The analysis shows -- quite unsurprisingly -- that there is the $BF$-theory fixed point $a=\lambda=0$, corresponding to the Gaussian theory, with $a$ being a relevant, $\lambda$ being an irrelevant coupling. Despite several different UV scenarios, all trajectories flow to the high temperature fixed point in the IR.

The fact that the flow remains virtually unchanged upon varying the $c_n$ suggests that one can arrive at good approximations for the flow by considering only few observables.

\section{Discussion and Outlook}

In this article, we have discussed the basic framework for background-independent renormalization in spin foam models. By carefully taking care of the embeddings of $2$-complexes in the space-time manifold, a clear notion of continuum limit arises, motivated by the construction of regular Borel measures on infinite-dimensional spaces.

The embedded $2$-complexes take on the meaning of scale, in the sense that they provide a cut-off for degrees of freedom, since they determine a finite-dimensional portion of infinite-dimensional state space. Also, since embedded $2$-complexes form a directed, partially ordered set, there is a natural hierarchy among the degrees of freedom, which does not use any notion of length scale.

The partial measures $\mu_\Gamma$ defined on each scale $\Gamma$ define the effective dynamics at that scale. The condition of cylindrical consistency ensures that the effective actions at different scales are compatible, and is the key ingredient to ensure the existence of the continuum measure $\mu$. The cylindrical consistency conditions are the background-independent analogue of the renormalization group flow equations. Indeed, if there is a background metric to associate lengths to the scales, the cylindrical consistency conditions indeed turn into Wilson's RG flow equations for effective actions.

It is important to note that the continuum theory does not consist of a limit in the usual sense. Rather, the continuum theory is equivalent to the collection of all effective theories at all scales.

Also, because one keeps track of embeddings, there is a clear action of space-time diffeomorphisms on the continuum theory. This allows for a natural notion of diffeomorphism-invariance of the path-integral measure.

From the examples we also have seen that, although the RG flow equations itself are background-independent, the solutions not necessarily are. Rather, some solutions spontaneously break diffeomorphism-symmetry by introducing a background solution. As expected, the group of diffeomorphisms then maps these solutions into each other, transforming the background structure accordingly.

For a solution to be background-independent, its effective theories need to satisfy very strong conditions, as given by (\ref{Eq:ConditionDiffInvariance}). This is advantageous for two reasons: Firstly, it allows to check for diffeomoprhism-invariance without knowing the solution in full generality (which will most likely not be possible in the case of quantum general relativity). Secondly, it strengthens the hope that the condition of diffeomorphism-invariance will severely restrict the RG flow, so that the non-renormalizability that plagues the perturbative ansatz might not affect the full, background-independent theory. This is certainly a point worthwhile to investigate further.

\paragraph*{Open questions:} One of the immediate limitations of the framework presented here is the condition of compactness of the gauge group $G$. This only allows for the treatment of quantum gravity with a Riemannian signature of the metric. Although there exist finite expressions for the vertex amplitude for the Lorentzian signature case, in which the gauge group is $G=SL(2,\mathbb{C})$ \cite{Engle:2008ev, Kaminski:2010qb}, the full path integral measure for one $2$-complex, does not, at this point, exist. In particular, it is not known how to make the sum over all intertwiners and spins finite, since for noncompact groups, there is no heat kernel regularization, and most likely no regularization at all, which does not break local gauge-invariance. Although there are finite versions of the full amplitude involving quantum deformations of $SL(2,\mathbb{C})$ \cite{Fairbairn:2011aa, Han:2010pz}, these do not fall into the framework presented in this article, due to the non-commutativity of the observables in those cases. Also, they are restricted to a positive cosmological constant.
    Note that the problem of defining a finite amplitude for Lorentzian signature is notorious for most other background-independent approaches to quantum gravity as well.

Additionally, the construction of spaces of generalized connections $\overline{\mathcal{A}}$ for non-compact groups is an unsolved problem until today. Also, cylindrical consistency alone will not guarantee the existence of the continuum measure, as the conditions for Kolmogorov's theorem are not automatically satisfied, as in the compact case \cite{Yamasaki:85}.

Even if all these problems might be overcome, it should be noted that, due to the fact that in the Lorentzian EPRL model all edges are fixed to be of a certain signature, in most cases time-like, such a model seems to essentially fix the causal structure from the outset. One would expect that the resulting theory only lets the conformal factor fluctuate, which would seem to miss the point of the geometric principles underlying the construction of the EPRL amplitude.

All of these points, as well as making contact with the implementation of RG flow in other background-independent approaches to quantum gravity \cite{Rivasseau:2011hm, Ambjorn:2014gsa}, should provide interesting research projects for the future.

\section*{Acknowledgements}

The author thanks Dr.~Bianca Dittrich and Dr.~Astrid Eichhorn for an invitation to Perimeter Institute, where the content of this article was presented as a talk. This work was funded by the project BA 4966/1-1 of the German Research Foundation (DFG).

\appendix

\section{Properties of $\overline{\mathcal{A}}$}\label{App:Properties}

In this section we go over the mathematical details of the construction of $\overline{\mathcal{A}}$ in more detail.

Recall that $\mathcal{M}$ is a semi-analytic, closed manifold. A two-complex $\Gamma$ is a finite collection of cells, i.e.~vertices, edges and faces, where each of those is, respectively, a $0$-, $1$- or $2$-dimensional, embedded semi-analytic manifold with the topology of, respectively, a point, a compact interval, or the bounded disc $D^1$. We demand the usual rules for cell complexes, i.e.~for every cell, its boundary is a collection of lower-dimensional cells, and for any two cells one of the three following cases is true: they do not intersect in $\mathcal{M}$, their intersection is a common subcell, or one is completely contained in the other's boundary.

Two $2$-complexes $\Gamma$ and $\Gamma'$ satisfy the relation $\Gamma\leq\Gamma'$, if and only if, for each $n$-cell $c$ of $\Gamma$ (with $n=0,1$ or $2$), there is a collection of $n$-cells $c'_1,\ldots, c'_k$ in $\Gamma'$ such that
\begin{eqnarray}
c\;=\;\bigcup_{i=1}^kc'_i.
\end{eqnarray}

\noindent Because we work in the semi-analytic category, for two $2$-complexes $\Gamma_{1,2}$ there is always a $2$-complex $\Gamma'$ finer than both, i.e.~$\Gamma_{1,2}\leq\Gamma'$. Its construction is straightforward by subdividing both $2$-complexes along their intersection, which are again semi-analytic submanifolds. \footnote{This is also true in the smooth category. However, in the smooth case, it can happen that the finer complex has infinitely many cells. On the level of $1$-complexes there is a -- technically very involved -- way around this \cite{Lewandowski:1999qr, Fleischhack:2000ij}, while whether a similar generalization of our framework to smooth, say CW, $2$-complexes exists or not is not clear at this point. So using the semi-analytic category for the moment is a technically convenient restriction, which might be overcome in the future. }

Note that with this relation, the set of $2$-complexes becomes a directed, partially ordered set.

An orientation $\omega$ of $\Gamma$ denotes an individual orientation for each edge $e$ and face $f$ in $\Gamma$ (vertices always implicitly carry the negative orientation). Denote the set of orientations $\Omega_\Gamma$, and the set of all pairs of edges $e$ and faces $f$ such that $e\subset f$ as $E\ltimes F$, and let $G$ be a compact Lie group, then the configuration space $\mathcal{A}_\Gamma$ is the set of all maps
\begin{eqnarray}
A:\;\Omega_\Gamma\times E\ltimes F\;\longrightarrow\;G
\end{eqnarray}

\noindent such that $A(\omega, (ef))=A(\tilde\omega,(ef))^{\pm1}$, depending on whether the orientations of the edge $e$ in $\omega$ and $\tilde\omega$ agree or disagree. Of course, upon a choice of orientation of $\Gamma$, $\mathcal{A}_\Gamma$ becomes naturally isomorphic to $G^{E\ltimes F}$, an isomorphism that we will use implicitly in what follows, as well as the rest of the article.\footnote{The reason for the unwieldy definition here is that oriented $2$-complexes do not, technically, form a partially ordered set. This is because differently oriented $\Gamma$, $\tilde\Gamma$ satisfy $\Gamma\leq\tilde\Gamma$ and $\tilde\Gamma\leq\Gamma$, but $\Gamma\neq\tilde\Gamma$. Nevertheless, in the rest of the article we will ignore this technicality, and assume that each $\Gamma$ carries an orientation.}

For $\Gamma\leq\Gamma'$, define the coarse graining map $\pi_{\Gamma'\Gamma}:\mathcal{A}_{\Gamma'}\to\mathcal{A}_\Gamma$ (for a chosen orientation on both $2$-complexes) as in (\ref{Eq:CoarseGrainingMap}) to be

\begin{eqnarray}\label{Eq:AppendixCoarseGrainingMap}
\pi_{\Gamma'\Gamma}\big(\{h_{e'f'}\}\big)_{ef}\;:=\;\overleftarrow{\prod_{e'\subset e,\,f'\subset f}}h_{e'f'}^{[e',e]},
\end{eqnarray}

\noindent where $[e',e]=\pm 1$ depending on whether the orientation on $e'$ agrees or disagrees with the one induced by the orientation of $e$. The ordered product is the product of group elements $h_{e'f'}$ going through all $e'\subset e$, starting at the beginning of $e$ until the end, according to its orientation.

It is an easy task to check that this definition is in fact independent of the choice of orientation on either $2$-complex, making $\pi_{\Gamma'\Gamma}$ well-defined, and satisfies $\pi_{\Gamma'\Gamma}\pi_{\Gamma''\Gamma'}=\pi_{\Gamma''\Gamma}$.\\[5pt]

Following \cite{Thiemann:2007zz}, we define the projective limit $\overline{\mathcal{A}}$ to be the subset of the Tychonoff product $\mathcal A_\infty\;=\;{\times}_\Gamma\mathcal{A}_\Gamma$ satisfying cylindrical consistency, i.e.
\begin{eqnarray}
\overline{\mathcal{A}}\;:=\;\big\{A\in\mathcal{A}_\infty\big|\Gamma\leq\Gamma'\;\Rightarrow\;
\pi_\Gamma(A)=\pi_{\Gamma'\Gamma}\,\pi_{\Gamma'}(A)\big\},
\end{eqnarray}

\noindent where $\pi_\Gamma:\mathcal{A}_\infty\to \mathcal{A}_\Gamma$ is the surjective projection onto the $\Gamma$th factor. Because all $\pi_\Gamma$ are continuous in the Tychonoff topology, and all $\pi_{\Gamma'\Gamma}$ are continuous as well, $\overline{\mathcal{A}}$ is a closed subset of $\mathcal{A}_\infty$, and because the latter is compact, so is the former, when we choose the subset topology, which we will do in what follows.

The continuous functions on $\overline{\mathcal{A}}$ contain the set of cylindrical functions
\begin{eqnarray}
{\rm Cyl}(\mathcal{A})\;=\;\big\{\mathcal{O}\in C^0(\overline{\mathcal{A}})\,\big|\,\textrm{there is }\Gamma,\;\mathcal{O}_\Gamma\in C^0(\mathcal{A}_\Gamma)\;:\;\mathcal{O}=\mathcal{O}_\Gamma\,\pi_\Gamma\big\}
\end{eqnarray}

\noindent as dense subset. For $\Gamma\leq\Gamma'$ one has the maps $\iota_{\Gamma\Gamma'}:C^0(\mathcal{A}_\Gamma)\to C^0(\mathcal{A}_{\Gamma'})$, defined by $\iota_{\Gamma\Gamma'}f:=f\,\pi_{\Gamma'\Gamma}$. Note that the embedding maps $\iota_{\Gamma\Gamma'}$ can be continued to isometries on the Hilbert spaces $\mathcal{H}_\Gamma:=L^2(\mathcal{A}_\Gamma, d\mu_{\rm Haar})$ via
\begin{eqnarray}
\iota_{\Gamma\Gamma'}\;:\;\mathcal{H}_\Gamma\;\longrightarrow\;\mathcal{H}_{\Gamma'}.
\end{eqnarray}

\noindent The collection $\{\mathcal{H}_\Gamma,\,\iota_{\Gamma\Gamma'}\}$ has a directed limit in the category of Hilbert spaces, coinciding with $L^2(\overline{\mathcal{A}},d\mu_{\rm AL})$, where $\mu_{\rm AL}$ is the Ashtekar-Lewandowski measure.\\[5pt]

Consider a collection $\mu_\Gamma$ of regular Borel measures on $\mathcal{A}_\Gamma$, i.e.~for each $\Gamma$ there is a positive linear map $\Lambda_\Gamma:C^0(\mathcal{A}_\Gamma)\to\mathbb{C}$ satisfying
\begin{eqnarray}
\Lambda_{\Gamma}(\mathcal{O}_\Gamma)\;=\;\int_{\mathcal{A}_\Gamma}d\mu_\Gamma\,\mathcal{O}_\Gamma.
\end{eqnarray}

\noindent Let furthermore the measures be normalized: $\Lambda_\Gamma(1)=1$, and satisfy
\begin{eqnarray}\label{Eq:AppendixCylindricalConsistency02}
(\pi_{\Gamma'\Gamma})_*\mu_{\Gamma'}\;=\;\mu_\Gamma.
\end{eqnarray}

\noindent Then there is a positive linear map $\Lambda:\textrm{Cyl}(\overline{\mathcal{A}})\to\mathbb{C}$ defined by $\Lambda(\mathcal{O})=\Lambda_\Gamma(\mathcal{O}_\Gamma)$ for $\mathcal{O}$ being cylindrical over $\Gamma$. Due to (\ref{Eq:AppendixCylindricalConsistency02}), $\Lambda$ is well-defined. Clearly, $\Lambda$ is bounded by $1$, so that it can be continued to $C^0(\overline{\mathcal{A}})$, so that there is a regular Borel measure $\mu$ on $\overline{\mathcal{A}}$. It is an elementary exercise to prove that the Haar measures $\mu_{\rm Haar}$ on each $\mathcal{A}_\Gamma$ satisfy the cylindrical consistency conditions (\ref{Eq:AppendixCylindricalConsistency02}). The resulting measure is precisely the Ashtekar-Lewandowski measure $\mu_{\rm AL}$.

\section{Connection to canonical framework}\label{App:Canonical}

The framework presented in this article needs only to be slightly amended in order to include manifolds with boundary. So let us assume that $\mathcal{M}$ is compact with closed boundary $\partial \mathcal{M}$. For $2$-complexes $\Gamma$ we allow only those which satisfy $\partial \Gamma:=\Gamma\cap \partial M\subset \Gamma^{(1)}$, i.e.~a face may intersect with the boundary only in some of its boundary edges. Thus $\partial\Gamma$ forms a graph.

Note however that we allow for an edge in the $\partial\Gamma$ to have more than one face in $\mathcal{M}$ meeting it. We will come back to this point later.

For each $2$-complex $\Gamma$ the configuration space is given by
\begin{eqnarray}
\mathcal{A}_\Gamma\;=\;G^{E\ltimes F}\times G^{\partial E},
\end{eqnarray}

\noindent where $\partial E$ denotes the number of edges in $\partial \Gamma$.

The ordering relation between $2$-complexes is equivalent to the case in which there is no boundary: $\Gamma\leq\Gamma'$ if and only if every cell in $\Gamma$ can be composed of cells in $\Gamma'$. Note that, because $\Gamma$ and $\Gamma'$ are both embedded in $\mathcal{M}$, this implies $\partial \Gamma$ to be a subgraph of $\partial \Gamma'$. For this we similarly write $\partial\Gamma\leq\partial\Gamma'$.

For $\Gamma\leq\Gamma'$ we define the coarse graining map $\pi_{\Gamma'\Gamma}:\mathcal{A}_{\Gamma'}\to\mathcal{A}_\Gamma$ to be
\begin{eqnarray}\label{Eq:AppendixCoarseGraining1}
\pi_{\Gamma'\Gamma}\big(\{h_{e'f'},g_{e'}\}\big)_{ef}\;:=\;\overleftarrow{\prod_{e'\subset e,\,f'\subset f}}h_{e'f'}^{[e',e]}\quad\textrm{for all }e\subset f\\[5pt]\label{Eq:AppendixCoarseGraining2}
\pi_{\Gamma'\Gamma}\big(\{h_{e'f'},g_{e'}\}\big)_{e}\;:=\;\overleftarrow{\prod_{e'\subset e}}g_{e}^{[e,e']}
\quad\textrm{for all }e\subset\partial\Gamma.
\end{eqnarray}

\noindent Unsurprisingly, $[e, e']=\pm1$ denotes the relative orientation of $e$ and $e'\subset e$.

A choice of a theory amounts to a choice of (regular Borel) measures $\mu_\Gamma$ on each  $\mathcal{A}_\Gamma$, satisfying the cylindrical consistency conditions
\begin{eqnarray}\label{Eq:AppendixCylindricalConsistency}
(\pi_{\Gamma'\Gamma})_*\mu_{\Gamma'}\;=\;\mu_\Gamma\qquad\textrm{whenever }\Gamma\leq\Gamma',
\end{eqnarray}

\noindent which is completely analogous to (\ref{Eq:CylindricalConsistency}). Notice, however, that Fubini's theorem allows us to integrate out all $h_{ef}$, without integrating out the $g_e$, leaving us, for each $\Gamma$, with a positive linear map $\eta_\Gamma\;:\;C^0(G^{\partial E})\;\to\;\mathbb{C}$, by
\begin{eqnarray}
\eta_\Gamma[\psi]\;:=\;\int_{G^{\partial E}}\left(\int_{G^{E\ltimes F}}d\mu_{\Gamma}(h_{ef}, g_e)\right)\,\psi(g_e).
\end{eqnarray}

\noindent Of course, this is just taking the expectation value $\langle\psi\rangle_\Gamma$, for $\psi$ interpreted as observable in $C^0(\mathcal{A}_\Gamma)$.

Cylindrical consistency (\ref{Eq:AppendixCylindricalConsistency}) has two immediate consequences:

\begin{itemize}
\item The linear map $\eta_\Gamma$ does \emph{not} depend on $\Gamma$, only on $\partial\Gamma$. To see this just note that for each $\Gamma_1$ and $\Gamma_2$ with $\partial\Gamma_1=\partial\Gamma_2$, there is a $\Gamma'$ with $\Gamma_{1,2}\leq\Gamma'$ and $\partial\Gamma'=\partial\Gamma_{1,2}$.

    We can therefore also write $\eta_{\gamma}$ to indicate that it just depends on the boundary graph $\gamma=\partial\Gamma$.
\item If there are two boundary graphs $\gamma\leq\gamma'$ with number of edges $E_\gamma\leq E_{\gamma'}$, then one has
\begin{eqnarray}\label{Eq:AppendixEta}
\eta_{\gamma}\;=\;\eta_{\gamma'}\;\iota_{\gamma\gamma'},
\end{eqnarray}
\noindent where the embedding map $\iota_{\gamma\gamma'}:C^0(G^{E_\gamma})\to C^0(G^{E_{\gamma'}})$ is the pull back of the projection maps $\pi_{\gamma'\gamma}$ on the boundary (\ref{Eq:AppendixCoarseGraining2}), i.e.~$\iota_{\gamma\gamma'}\psi=\psi\,\pi_{\gamma'\gamma}$.
\end{itemize}

With the boundary graphs $\gamma$ and the respective configuration spaces $\mathcal{A}_\gamma=G^{E_\gamma}$, forming a directed set as well, one can form the boundary configuration space $\overline{\mathcal{A}}_{\partial\mathcal{M}}$, which is a standard construction in loop quantum gravity \cite{Thiemann:2007zz, Ashtekar:1994mh,Ashtekar:1995zh}, going along the same lines as the construction of $\overline{\mathcal{A}}$ in chapter \ref{Sec:HolonomySFM}.

Because of (\ref{Eq:AppendixEta}), there is a positive linear map $\eta:C^0(\overline{\mathcal{A}}_{\partial\mathcal{M}})\,\to\,\mathbb{C}$ which satisfies $\pi_\gamma^*\eta\;=\;\eta_\gamma$.

\paragraph*{Examples:}
Assume we have a manifold $\mathcal{A}$ with $\partial{M}\simeq\Sigma$ a compact, connected $3$-manifold, interpreted as ``space''. Then $\eta$ defines a regular Borel measure on $\overline{\mathcal{A}}_{\partial\mathcal{M}}$, hence a linear functional on a dense subset of the kinematical boundary Hilbert space $\mathcal{H}=L^2(\overline{\mathcal{A}}_{\partial\mathcal{M}},d\mu_{\rm AL})$, the states of which have an interpretation as quantized $3$-geometries \cite{Thiemann:2007zz, Livine:2007vk, Freidel:2010aq}. Thus, $\eta$ defines a generalized boundary state, which plays the role of Everett's ``wave function of the universe'' \cite{EverettThesis, Hartle:1983ai}.\\[5pt]

\begin{figure}[t!]
\centering
\def\svgscale{0.5}
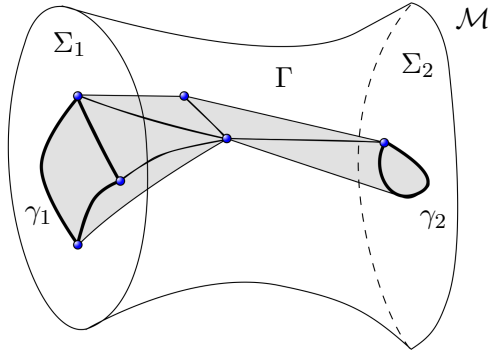
\caption{Manifold $\mathcal{M}$ with boundary $\partial\mathcal{M}=\Sigma_1\sqcup\Sigma_2$.
    }\label{Fig:Figure12}
\end{figure}

Another case is that of a manifold $\mathcal{M}$ with topology $\mathcal{M}\simeq\Sigma\times [0,1]$, $\Sigma$ again playing the role of ``space'' as a Cauchy hypersurface, as in figure \ref{Fig:Figure12}. Then the boundary is $\partial\mathcal{M}\simeq\Sigma\sqcup\Sigma$, and $\eta$ is a positive linear form on $C^0(\overline{\mathcal{A}}_{\partial \mathcal{M}})\simeq C^0(\overline{\mathcal{A}}_{\Sigma})\otimes C^0(\overline{\mathcal{A}}_{\Sigma})$, which can in turn be used to define a sesquilinear form on a dense subset of the kinematical boundary Hilbert space via
\begin{eqnarray}
\langle\psi|\phi\rangle_{\rm phys}\;:=\;\eta[\overline{\psi}\otimes \phi]
\end{eqnarray}

\noindent with $\psi,\phi\in C^0(\overline{\mathcal{A}}_{\Sigma})$. Here the ``complex conjugate'' of a state is the one obtained by reversing the orientation of all edges on the boundary (but not of the vertices). This defines a rigging map $\eta:C^0(\overline{\mathcal{A}}_{\Sigma})\to \big(C^0(\overline{\mathcal{A}}_{\Sigma})\big)'$ in the sense of refined algebraic quantization \cite{Marolf:1995cn, Giulini:1998rk}, and if the continuum path integral measure is such that $\eta[\overline{\psi}\otimes \phi]=\overline{\eta[\overline{\phi}\otimes \psi]}$, then this defines a physical inner product. Note that this is automatically the case if $\mu$ is invariant under all diffeomorphisms of $\mathcal{M}$, since these include the ones switching the two boundaries.

If should be noted that if $\mu$ is invariant under $\textrm{Diff}^{\omega/2}(\mathcal{M})$, then $\eta$ is invariant \emph{only} under those diffeomorphisms of the boundary which can be extended to a diffeomorphism on all of $\mathcal{M}$. In particular, the physical inner product might not necessarily be invariant under large diffeomorphisms of $\Sigma$, only under those which are in the connected component of the identity. This in particular suggests that the elements of $\textrm{Diff}^{\omega/2}(\Sigma)/\textrm{Diff}_0^{\omega/2}(\Sigma)$ should act as unitary operators on the kinematical Hilbert space, but not as gauge transformations (see e.g~the discussion in \cite{Giulini:2006na}).

\paragraph*{General remarks:} We should note that it is strictly necessary in this approach to allow for several faces to meet in one boundary edge, in order to have the group $\textrm{Diff}^{\omega/2}(\mathcal{M})$ act on the space of $2$-complexes \emph{and} the $2$-complexes forming a partially ordered set. So allowing  for more than one-valent edges on the boundary is crucial to determine the action of the diffeomorphism group on the bulk and the boundary Hilbert space the correct way.

For those two-complexes $\Gamma$ in which all boundary edges $e$ are part of the boundary of precisely one face $f$, the condition of gauge-invariance, which is a sensible requirement for the measures $\mu_\Gamma$, but not strictly necessary, will severely restrict the interplay between $g_e$ and $h_{ef}$. In most spin foam models, these two will actually be set equal, i.e.~the measure will necessarily contain a term of the form $\delta(h_{ef},g_e)$, see e.g.~\cite{Kaminski:2009fm, Bahr:2010bs, Bahr:2011aa}.

Also, one might want to construct the measures $\mu_\Gamma$ in such a way, that some kind of cobordism functoriality as in topological field theories should hold, i.e.~for manifolds $\mathcal{M}_1$, $\mathcal{M}_2$ glued together along some common boundary to satisfy
\begin{eqnarray}\label{Eq:AppendixFunctoriality}
\eta_{\mathcal{M}_1}\eta_{\mathcal{M}_2}\,=\,\eta_{\mathcal{M}_1\#\mathcal{M}_2}.
\end{eqnarray}

\noindent It should be noted that an equation like (\ref{Eq:AppendixFunctoriality}) will only hold in this strong form for cases in which the group $G$ is finite -- and hence the boundary Hilbert space associated to one graph finite dimensional. Indeed, then one has arrived at the axioms for a TQFT in the sense of Atiyah \cite{Atiyah:1989vu}, which have the finite-dimensionality of their physical Hilbert spaces as an a posteriori consequence. In general, $\eta$ will be a sesquilinear form rather than an operator, for which an equation of the type of (\ref{Eq:AppendixFunctoriality}) can not even be formulated.


\end{document}